\documentstyle[aps,multicol]{revtex}         
\begin {document}

\title{Nonstationary Transport by Internal White Noise and 
"Localization" in Ratchets}

\author{ Nikolai Brilliantov$^{1,2}$ and Peter Strizhak $^{1,3}$} 
\address{\small\it{$^1$ Department of Chemistry, University of 
Toronto, Toronto, Canada M5S 3H6 \\
$^2$ Physics Department, Moscow State University,  Moscow 119899,Russia\\ 
$^3$ Institute of Physical Chemistry, National Academy of Sciences of Ukraine,
pr.Nauki, 31, Kiev 252038, Ukraine}}

\maketitle

\bigskip
\begin{abstract}

We revealed a right-left asymmetry of the inter-well mean first passage 
times for the Brownian particles in a ratchet potential under internal 
white noise. We showed analytically and numerically that this asymmetry 
gives rise to the following phenomena: (i)~nonstationary transport;
(ii)~localization or ultraslow (Sinai) diffusion for ratchets with a  
disorder in the asymmetry parameter of the potential;
(iii)~equilibrium "drift of labels" in circular ratchets.

\medskip\noindent PACS numbers: 82.20.Mj, 05.40.+j 
\end{abstract}

\begin{multicols}{2}          

It was demonstrated recently, that nonequilibrium fluctuations may support a 
stationary flux of the Brownian particles in a periodic potential with broken 
symmetry (ratchet). 
This fluctuation-induced transport has been of growing interest after 
Magnasco suggestion to apply a toy 
"ratchet and pawl" Feynman's model to riddle  a "mystery" of the 
directed transport in cells ~\cite{magnas1}.  
In subsequent studies ~\cite{bier1,doer1,magnas2,milon1,milon2,bierouss}
, some related models were proposed to explain the 
nature of molecular motors ~\cite{bier1,magnas2}, to develop a novel mass 
separation technique ~\cite{bierouss}, to design the Brownian engine with a 
negative resistance ~\cite{magnas3}, or to speculate how the ion channel 
junctions operate ~\cite{milon2}. According to 
the second law, no stationary transport occurs  under 
equilibrium fluctuations. To obtain a stationary motion, one has to break 
somehow the detailed balance. A possible way is to impose an 
additional colored noise  ~\cite{magnas1,bier1,doer1,hang1,dyk,hang2}.
One can also couple the system with nonequilibrium bath ~\cite{milon1}, 
with external (e.g. chemical) degrees  of freedom 
~\cite{bier1,magnas2,magnas3}, or with a deterministic chaotic process
~\cite{hondo}.

In the present study we consider the motion of the Brownian particles  in a 
ratchet potential (Fig.1a) under an equilibrium white noise. 
We found a right-left asymmetry for the mean first passage times (MFPT)
of the interwell transitions. This property of the MFPT 
entails the following phenomena: (i)~nonstationary transport of the 
Brownian particles; (ii)~ultraslow (Sinai) diffusion  
or "localization" in ratchets with a disorder (Fig.1b.); (iii)~equilibrium 
"drift  of labels" in  circular ratchets. 
The directed nonstationary transport of the ensemble of the Brownian particles 
occurs as its  peculiar route towards equilibrium ~\cite{remark1}. 
If the process is interrupted at some stage (e.g. removing a 
particle) a net transport is obtained. This scenario 
of the nonstationary transport seems to be realistic for biological systems. 

Nothing is perfectly periodic in nature, and any periodic structure has 
imperfections and distortions. Sometimes even weak disorder 
drastically changes a kinetic behavior of the system ~\cite{bouch}. If the 
ratchet potential is symmetric in average, but the asymmetry parameter,
$\alpha=L_{-}/L_{0}$ is stochastically distributed around 
$\alpha=1/2$ (Fig.1), the mean displacement of the particles is zero. 
Surprisingly, the mean square displacement behaves as 
$\sim \left[ \log(t) \right]^4 $ at large times $t$, contrary to $\sim t$ 
dependence for the usual diffusion. At $t \gg1$ 
$\left[ \log(t) \right]^4 /t \to 0$, so the particle is, in fact, localized 
on the length scale of the ordinary diffusion. The 
"localization" may be effectively used in biological systems, or in 
technology, instead of trapping molecules  in a deep potential well 
by forming a chemical bond. It costs much less energy to release the 
particle localized by the disordered ratchets, rather then waste energy, breaking 
the chemical bond. 

The directed nonstationary transport and ultraslow diffusion concern the 
Brownian motion on the infinite line. The other interesting 
phenomenon occurs in circular ratchets. From  one hand, the equilibrium flux 
in a ring is zero, from the other,  the interwell MFPT for transitions 
to the right is not equal to the MFPT of the left transitions. As a result 
a "drift of labels" may be detected, provided that the particles may be 
labeled and their paths may be traced. 
The "drift of labels" is not a thermodynamical flux and and the 
detailed balance at equilibrium is preserved.

To address these problems we consider the Brownian particle in a periodic 
ratchet potential (Fig.1a), subjected to the white noise, 
$\xi(t)$, $< \xi(t) \xi (t') >=2D \delta (t-t')$. Here 
$D=kT/\gamma$ is the diffusion, and $\gamma$ is the 
friction coefficient. The Langevin equation for the overdamped case reads:
\begin{equation}
\dot{x} = f(x)+\xi(t)
\label{lang}
\end{equation}
with  $f(x) =-\gamma^{-1} \partial U /\partial x$, and $~U(x)$ being  the 
ratchet potential. The corresponding Fokker-Plank equation is 
\begin{equation}
\partial_t {P} =-\partial_x \left( f P \right)+ 
D \partial_{xx} P
\label{FP}
\end{equation}
The Brownian motion in the piecewise linear potential may be effectively 
studied in terms of 
the MFPT ~\cite{bierdoer}. The escape time from a starting 
point, $x_0$, in the well is given as a solution of the equation 
~\cite{gard}:
\begin{equation}
D\frac{d^2T(x_0)}{dx_0^2}-\gamma^{-1} \frac{dU}{dx_0}\frac{dT(x_0)}{dx_0}+1=0
\label{time}
\end{equation}
The final point is either $L_+$, if the particle escapes to the right, 
or $-L_-$, if it overcomes the barrier in the opposite direction. The 
averaging of the escape time over $x_0$  gives $T_+$ and $T_-$, the MFPT 
for the right and for the left transitions respectively. 
To find $T_+$,  we use the continuity of 
$T'(x_0)$, and boundary conditions $T(L_+)=0$, $T'(-L_-)=0$, which are 
self-consistent at $\tau_{\pm}/T_{\pm} \ll 1$. 
Here $\tau_{\pm}=\gamma L_{\pm}/kTa_{\pm}$ are the {\it intrawell} relaxation 
times, and $a_{\pm}={U_0}/kT{L_{\pm}}$ characterize the force on the slopes.
The boundary conditions are reversed to obtain  $T_-$. 
We consider the case when  $\exp(U_0/kT) \gg 1$ and 
$\tau_{\pm}/T_{\pm} \ll 1$, i.e., the particle has enough
time to thermalize inside the well before it jumps over the 
barrier. Straightforward  analysis 
shows that the solution of Eq.(\ref{time}) is a sum of few terms 
~\cite{remark3}. The leading term is independent on $x_0$, and is of 
the order of $\exp(U_0/kT) \gg 1$. The other terms ~\cite{remark3}  
may be  neglected 
everywhere exept the regions in the very vicinity of the tops. 
If the distribution inside the well is sharply peaked around 
the bottom, these regions do not noticeably contribute to the MFPT. 
Particularly, this is true for the equilibrium distribution if 
$U_0/kT \ge 0.1$. Finally, we obtain  

\begin{equation}
T_{\pm}= a_{\pm}^{-1}{e^{U_0/kT}}/{Da_0}
\label{Tav}
\end{equation}
where $a_{0}^{-1}=a_{+}^{-1}+a_{-}^{-1}$.
This equation shows that $T_+ \neq T_-$ for the potential
with asymmetry. It is worth noting that the left-right asymmetry of 
the MFPT for the ratchet under the white noise occurs as the
preexponential factor. For the colored noise this factor is 
overshadowed by the asymmetry that appears in the exponent ~\cite{dyk}. 
The left-right asymmetry of the MFPT described by Eq.(\ref{Tav}) 
was confirmed in our numerical studies performed for the equilibrium 
intrawell distribution. To get the required white noise process (which is 
effectively  colored due to the finite time step) we decreased the time 
step until the results became insensitive to it further decrease 
in a range of two decades (the time step varied from 
$2\,10^{-4}$ to $10^{-6}$, and was roughly two orders of magnitude 
smaller, then that, used in ~\cite{doerstat}). We found that the simplest Euler 
method (see e.g. ~\cite{doerstat}) is quite effective for the piecewise 
linear potential.

As a consequence of the left-right asymmetry in the MFPT, the directed 
motion appears before the system reaches the stationary probability 
distribution, $P_{eq}(x) \propto \exp (-U(x)/kT)$, which is a current free, 
if no density gradients present in the system. Thus, any initially localized 
distribution, say $P(x,t;x_0,t_0)=\delta (x-x_0)(t-t_0)$, evolves to the 
$P_{eq}(x)$, defined on the ratchet, with left and right 
boundary  $-L_{\infty}$, $L_{\infty} \gg L_0$. 
In the sequence of the time scales, that characterize the Brownian 
motion in the overdumped case, the shortest time scale corresponds to
the intrawell relaxation time $\tau_{\pm}$. The next time 
scale is related to the interwells transition times, $T_{\pm}$. Finally, 
$T_{\infty} \sim (L_{\infty}/ L_0)T_{\pm}$  defines the global relaxation 
time. Under condition $\exp(U_0/kT)\gg 1 $, the following time-scales 
hierarchy holds,  $\tau_{\pm} \ll T_{\pm} \ll T_{\infty}$, and the 
nonstationary transport takes place for $T_{\pm} \ll t \ll T_{\infty}$.
To describe the nonstationary 
transport we introduce the population  of the $k$-th well as 
$n_k(t)=\int_{L_0k}^{L_0(k+1)}P(x,t)dx$, and write the kinetic equation:
\begin{equation}
\dot{n}_k=W_{-}n_{k+1}+W_{+}n_{k-1}-\left(W_{-}+W_{+} \right)n_{k}
\label{nd}
\end{equation}
where $W_{\pm}$ give transition rates, which may be estimated as 
$W_{\pm} \simeq T_{\pm}^{-1}$ ~\cite{remark2}. 
Performing transformation from $n_k$ in the discrete space to $n(x)$ in 
continuous space coordinate, we obtain the coarse-grained equation:

\begin{equation}
\partial_{t}n(x,t)\simeq -V \partial_{x}n(x,t) + \tilde{D}\partial_{xx}n(x,t)
\label{nx}
\end{equation}
with $V=L_0(T_{+}^{-1}-T_{-}^{-1})$ and 
$\tilde{D}=\frac{1}{2}L_0^2(T_{+}^{-1}+T_{-}^{-1})$. More rigorous derivation 
of the coarse-grained equation, based on the separation of different 
time-scales  ~\cite{brilk} yields the same result 
 for $t \gg T_{\pm}$. 

Eq.(\ref{nx}) shows that the ensemble of Brownian particles moves with a 
constant velocity $V$, which increases with enhancing asymmetry of 
the potential as 
$V=(U_0/kT)^2e^{-(U_0/kT)}(D/L_0)(\frac{1}{1-\alpha}-\frac{1}{\alpha})$ 
(Fig.2). The increase of the distribution width with time is 
described by the renormalized diffusion coefficient $\tilde{D}$.

We apply the coarse-grained description for the case of the 
external field $U_{ex}(x)=-Ex$. Calculations of the MFPT for the small fields, 
$EL_0/U_0 \ll1$, yield Eq.(\ref{Tav}), with $a_{\pm}$ substituted by 
$a_{\pm}\pm E/kT$, and $U_0/kT$ 
substituted by $(U_0\pm EL_{\pm})/kT$. In the linear approximation for $E$,
the average velocity reads:
\begin{equation}
V_{E}=V+(\tilde{D}/kT)E
\label{V}
\end{equation}
where $V$ and $\tilde{D}$ are the field-free values. 
The top insert in Fig.2 depicts dependence of the average velocity upon 
the external force.  The particles move  uphill, if the 
field does not exceed the critical value, $E_c=(2kT/L_0)(1-2\alpha)$.

The kinetic behavior of the system in the vicinity of the critical field, i.e. 
near $V_E=0$ drastically changes under (even weak) disorder.
The disorder may be either in the values of $E$, or in the values of 
$T_{\pm}$, i.e. in the asymmetry parameter $\alpha$. For such a 
"quenched disorder" ~\cite{bouch}, one has stochastically distributed 
transition probability rates,  $W_{+} \simeq T_{+}^{-1}$ and 
$W_{-}\simeq T_{-}^{-1}$ (probabilities for the right and for the left "jumps"). 
If $\langle \log (W_{+})\rangle=\langle \log (W_{-})\rangle$, 
the average displacement of the particle is zero, while the average square 
displacement behaves asymptotically as  $\sim (\log t)^4$ 
~\cite{bouch,kehr}. This  behavior is called as ultraslow (Sinai) 
diffusion, and may be regarded as a localization 
in the length scale of the ordinary diffusion~\cite{bouch}. 
If the external field is zero, $E=0$, the 
ultraslow diffusion occurs in ratchets, which are symmetric in average.

To prove the existence of the ultraslow (Sinai) diffusion, we applied an 
approach commonly used to analyze this phenomenon ~\cite{kehr}.
Namely, we consider a motion of a particle in a segment of $N=20$ ratchets 
with $\alpha$ distributed uniformly between $0.2$ and $0.8$, so that the 
average length of the left slope was equal to the average length of the 
right slope. The reflecting left boundary and adsorbing right boundary were 
used. We 
calculated numerically the distribution $P(T_N)$ of times $T_N$, 
when the particle, initially located at the left boundary, reached  the 
adsorbing boundary. For the ordinary $1d$ diffusion, the $T_N \gg 1$ 
asymptotics reads  $P(T_N) \sim T_N^{-3/2}$, while for the 
Sinai diffusion $P(T_N) \sim T_N^{-1}$ ~\cite{kehr}. The numerical results 
shown in Fig.3 illustrate that the ultraslow diffusion does take place in 
ratchets with the disorder. 

The transition probability rates $W_{\pm}$, 
as well as the kinetic equation (~\ref{nx}) refer to the time 
coarse-grained description and may be unambiguously applied only 
for $t \gg T_{\pm}$.  To analyze the detailed balance or the thermodynamic
flux $J(x)=\lim_{\Delta t \to 0} (N_+-N_-)/\Delta t$, one 
needs more refined approach, since $\Delta t \ll T_{\pm}$. Here 
$N_+$ and $N_-$ denotes correspondingly the  number of particles that 
have crossed the plane located at $x$ in positive and negative directions during 
the time interval $\Delta t$. To show that the asymmetry of 
the $T_{\pm}(x)$ does not imply the non-zero equilibrium flux, we write 
for the flux, e.g. at $x=L_+$ (see Fig.1):
$J(x,t)=(1/\Delta t)[ \int_{-L_-}^{L_+}P(x,t)~\theta(\Delta t-T_+(x))dx-
\\ \int_{L_+}^{L_++L_0}P(x,t)~\theta(\Delta t-T_-(x))dx ]$, where 
the unit step function $\theta(t)$ guarantees that particles have crossed 
the plane within the $\Delta t$. Since $\Delta t \to 0$, we 
expand $T_{\pm}(x)$ ~\cite{remark3} near $x=L_+$:
$T_{\pm}(x)= \mid L_+-x \mid e^{U_0/kT}/a_0D$.  As a result, 
a zero flux is obtained for the equilibrium probability density $P_{eq}(x)$.

In spite of a lack of the thermodynamic current at equilibrium, very 
unusual equilibrium fluctuations may be observed in circular ratchets. 
If particles may be labeled and if their paths may be traced, 
 one observes that 
the average number of times that particle crosses some 
cross-section  clockwise, $\langle S_+  \rangle$, is not equal at 
equilibrium to the average number of the counter-clockwise crossings, 
$\langle S_-  \rangle$. Therefore, clockwise (or counter-clockwise) 
"drift of labels" takes place in equilibrium.  Obviously, this is not a 
thermodynamical current, and no useful work may be 
extracted. We  quantify the "drift of labels" by 
$S=\left( \langle S_+ \rangle /\langle S_- \rangle -1\right)$.
We also introduce the mean times of the clockwise turn over 
the ring, $T_{C +}$, and of the counter-clockwise turn, $T_{C -}$. 
If, again, $\exp (U_0/kT) \gg 1$ and $\tau_{\pm} /T_{\pm} \ll 1$, a 
straightforward solution of Eq.(\ref{time}) gives for the ring of 
$M$ ratchets:

\begin{equation}
T_{C \pm}=\frac{e^{U_0/kT}}{Da_0} \left( \frac{M(M-1)}{2a_0}+
\frac{M}{a_{\pm}} \right)
\label{circ}
\end{equation}
and we estimate $S$ as 
$(T_{C +}^{-1} /T_{C -}^{-1}-1)=\frac{4\alpha-2}{(M+1)-2\alpha}$. 
The "flux of labels" disappears at large $M$, and thus it is an 
intrinsically microscopic phenomenon. Our numerical studies of the 
Brownian motion on the ring, of $M=4$ ratchets with $\alpha=0.8$ 
confirmed that the average current ceased to zero, while $S$ 
relaxed to $0.29$, which is in a good agreement with the theoretical
estimate $S=0.35$.

In conclusion, we studied the Brownian motion of particles in a ratchet 
potential under the internal white noise. We found that the 
inter-well mean-first passage times have a right-left asymmetry for 
the equilibrium and near-equilibrium intrawell distributions of particles. 
We have shown  analytically and 
numerically that this asymmetry gives rise to the following effects: 
(i)~nonstationary transport; (ii)~localization or ultraslow (Sinai) 
diffusion under disorder in the parameter of asymmetry of the ratchet 
potential; (iii)~unusual equilibrium fluctuations in 
circular ratchets, which may be called as a "drift of labels".

We are indebted to M.~Bier, R.~Kapral, and A.~Malevanets for valuable comments 
and discussions.

\bigskip\bigskip

Fig.~1 Ratchet potential,  $L_0=L_++L_-$ (a); ratchet with a disorder 
of the asymmetry parameter (b).

Fig.~2 The average displacement versus time for the ensemble, initially 
localized at $x=0$. Bottom insert: the average velocity, $V$, versus the 
asymmetry parameter, $\alpha$. Top insert: $V$  versus the external force, 
$E$. The dots - simulations, the lines - theory; $kT=0.085$, $\alpha=0.8$,
$\gamma =1$, $L_0=1$, $U_0=1$.

Fig.~3 The distribution of escape times for the segment of 
20 ratchets with reflecting left boundary. Bottom curve - the 
symmetric ratchet, $\alpha=1/2$. Upper curve - the ratchet with a disorder;
a flat distribution of $\alpha $ was used. 

\end{multicols}     

\begin{references}
\bibitem{magnas1} M.O.Magnasco, Phys.Rev.Lett. {\bf 71}, 1447 (1993).
\bibitem{bier1} R.D.Astumian and M.Bier, Phys.Rev.Lett. {\bf 72}, 1766 (1994).
\bibitem{doer1} C.R.~Doering, W.~Horsthemke  and J.~Riordan, 
Phys.~Rev.~Lett. {\bf 72}, 2984 (1994).
\bibitem{magnas2} M.O.Magnasco, Phys.Rev.Lett. {\bf 72}, 2656 (1994).
\bibitem{milon1} M.M.Millonas, Phys.Rev.Lett. {\bf 74} 10, (1995).
\bibitem{milon2} M.M.Millonas and  D.R.Chialvo, Phys.Rev.Lett. {\bf 76},
550 (1996).
\bibitem{bierouss} M.Bier, and R.D.Astumian, Phys.Rev.Lett. {\bf 76}, 4277 
(1996); M.Bier, Phys. Lett. A {\bf 211}, 12 (1996); J.Rousselet, L.Salome, 
A.Ajdari, and J.Prost, Nature(London) {\bf 370}, 446 (1994); J.Prost, J.F.Chauwin, 
L.Peliti, and A.Ajdari,  Phys.Rev.Lett. {\bf 72}, 2652 (1994)
\bibitem{magnas3} G.A.Cecchi and M.O.Magnasco, Phys.Rev.Lett. {\bf 76},
1968 (1996).
\bibitem{hang1}R.Bartussek, P.Reimann, and P.Hanggi, 
Phys.Rev.Lett. {\bf 76}, 1166 (1994)
\bibitem{dyk} M.M.Millonas, and M.I.Dykman, Phys.Lett. {\bf A185}, 65 (1994).
\bibitem{hang2} J.Luczka, R.Bartussek, and P.Hanggi, Europhys. Lett.
{\bf 31} 431 (1995); P.Jung, J.G.Kissner, and P.Hanggi, Phys.Rev.Lett.
 {\bf 76}, 3436 (1996).
\bibitem{hondo} T.Hondo and Y.Sawada Phys.Rev.Lett. {\bf 75} 3269 (1995).
\bibitem{remark1}Note, that the ensemble of the Brownian particles,
 being in equilibrium with  the bath by their impulses, is not 
necessarily  in a configurational equilibrium.
\bibitem{bouch}J.P.Bouchaud and A.Georges, Phys.Reports,  {\bf 195}, 
127 (1990).
\bibitem{bierdoer}M.Bier and R.D.Astumian, Phys.Rev.Lett. {\bf 71}, 1649 
(1993);  C.R.Doering and J.C.Gadoua, Phys.Rev.Lett. {\bf 69}, 2318 (1992).
\bibitem{gard}C.W.Gardiner, {\it Handbook of Stochastic Methods}
(Springer, Berlin, 1985), 2nd ed.
\bibitem{remark3}For $T_{+}(x)$ at $x>0$ one has:
$T_{+}(x>0)=(e^{U_0/kT}-e^{a_+x/kT})/Da_+a_0+(e^{-U_0/kT}e^{a_+x/kT}-1)/Da_+a_-$, 
with  similar expressions for $T_{+}(x<0)$, and for $T_{-}(x)$. 
\bibitem{doerstat}T.S.Elston, and  C.R.Doering, J.Stat.Phys., {\bf 83}, 
359 (1996).
\bibitem{remark2}We found that  the interwell transition rates were close 
to the reverse values of the  MFPT for $U_0/kT \ge 0.1$. 
\bibitem{brilk}N.V.Brilliantov, and A.I.Kviatkievitch, 
Sov. Phys. Solid State,   v.31, 2060 (1989).
\bibitem{kehr}K.W.Kehr and K.P.N.Murthy, Phys.Rev. {\bf A40}, 5728 (1990);
K.P.N.Murthy, S.Rajasekar, and K.W.Kehr, J.Phys.A:Math.Gen. {\bf 27}, 
L107 (1994).
\end{references}
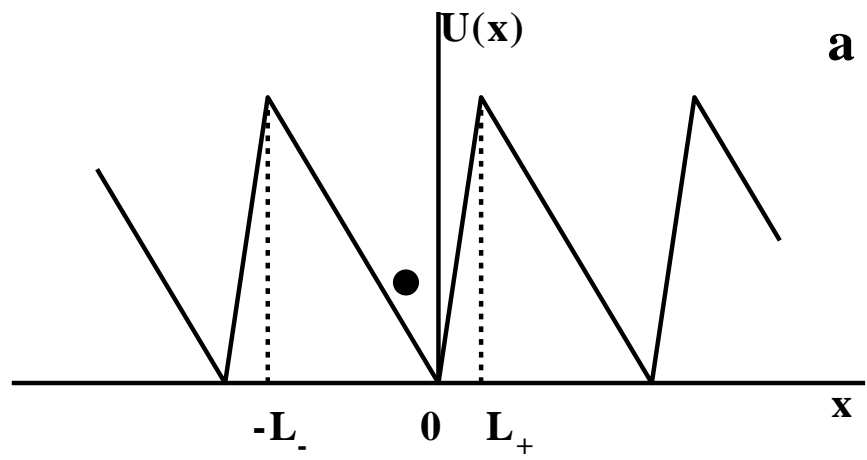
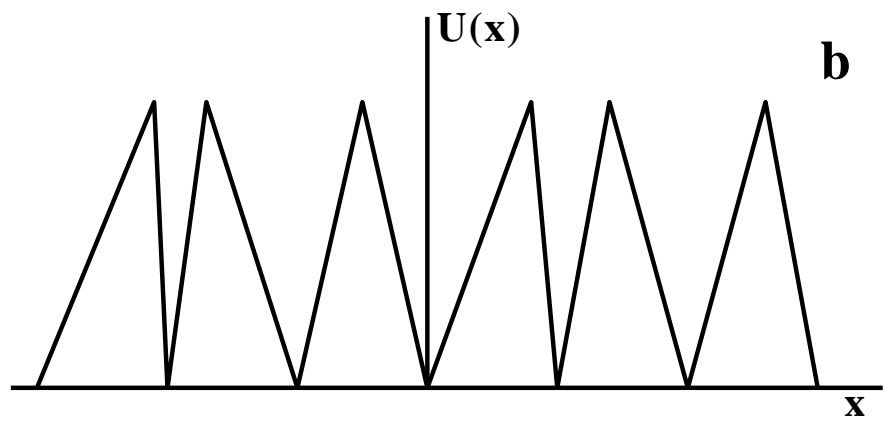
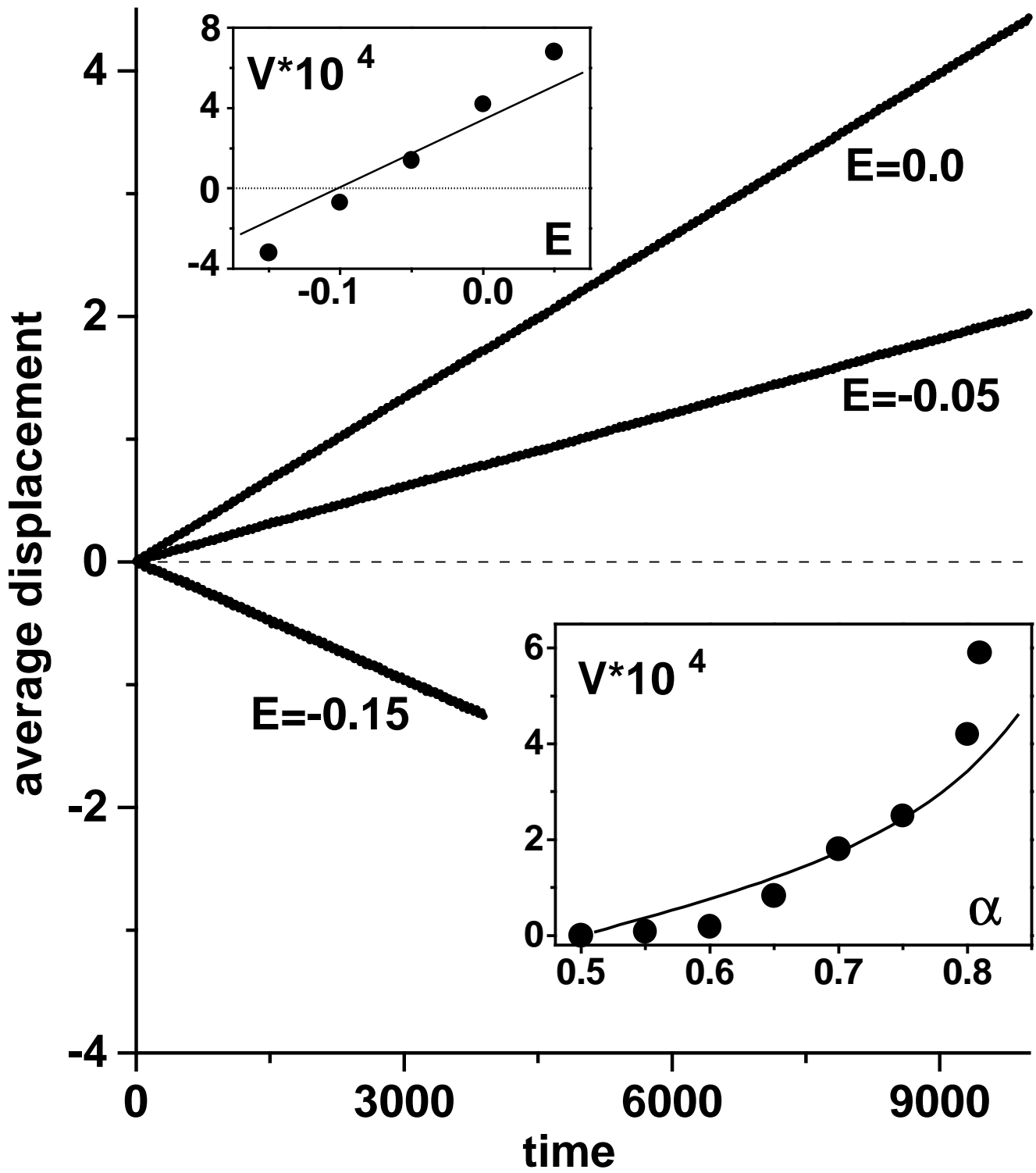
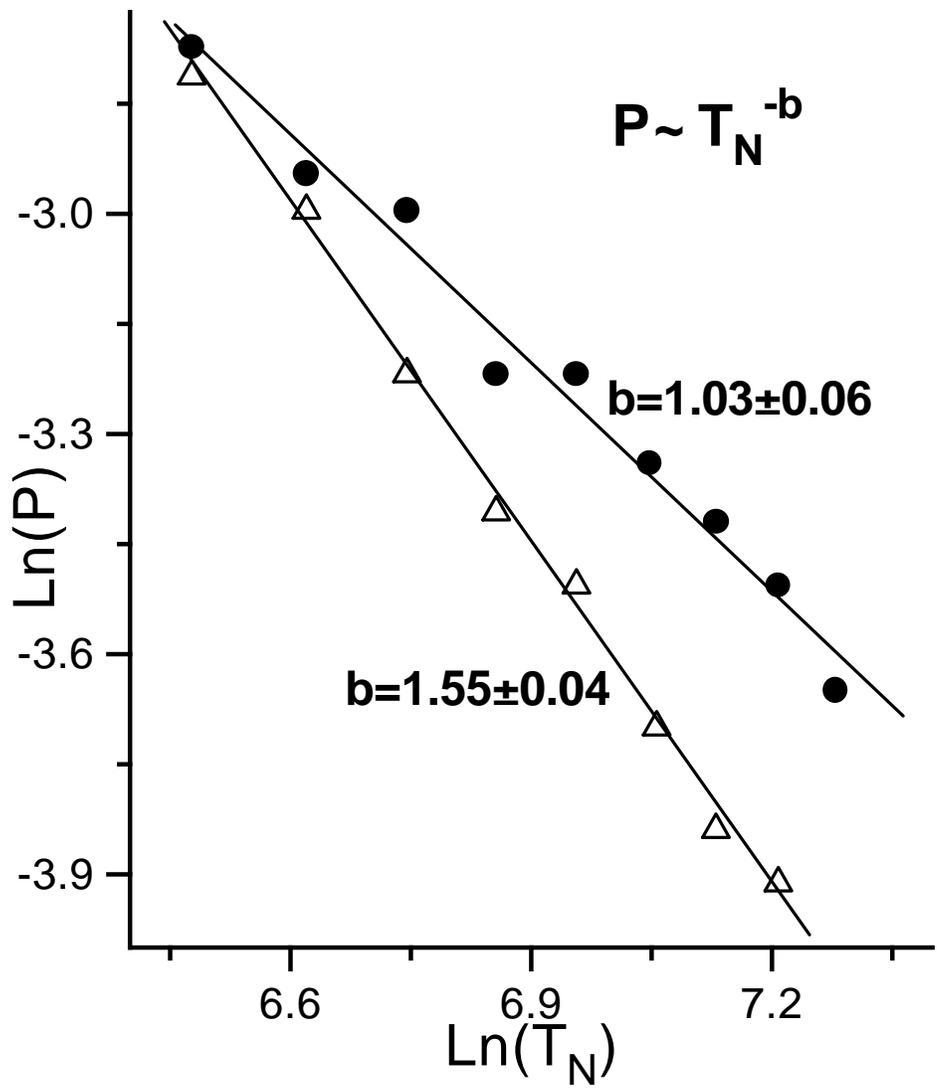
\end{document}